\pdfoutput=1
\documentclass[12pt]{iopart}
\usepackage{upgreek}
\usepackage{graphicx,float}

\begin{document}

\title{Effect of heat treatment on mechanical dissipation in Ta$_2$O$_5$ coatings}

\author{I W Martin$^1$, R  Bassiri$^1$, R Nawrodt$^{1,2}$, M  M  Fejer$^3$,\\ A  Gretarsson$^4$, E Gustafson$^5$, G  Harry$^6$, J Hough$^1$,\\ I
MacLaren$^1$, S Penn$^7$, S Reid$^1$, R Route$^3$, S Rowan$^1$,\\ C Schwarz$^2$, P Seidel$^2$, J Scott$^1$ and A L Woodcraft$^{8,9}$}

\address{$^1$ SUPA\footnotemark[1], Department of Physics and Astronomy, University of Glasgow, Glasgow, G12 8QQ, Scotland.}
\address{$^2$Institute of Solid State Physics, University of Jena, Helmholtzweg 5, D-07743 Jena, Germany.}
\address{$^3$ Edward L Ginzton Laboratory, Stanford University, Stanford, CA 94305-4088, USA.}
\address{$^4$ Embry-Riddle Aeronautical University, Prescott, AZ 86301, USA.}
\address{$^5$ LIGO Laboratory, California Institute of Technology 18-34, Pasadena, CA 91125, USA.}
\address{$^6$ LIGO Laboratory, Massachusetts Institute of Technology, Cambridge, MA 02139, USA.}
\address{$^7$ Department of Physics, Hobart and William Smith Colleges, Geneva, NY 14456, USA.}
\address{$^8$ SUPA\footnotemark[1], Institute for Astronomy, University of Edinburgh, Blackford Hill, Edinburgh EH9 3HJ, Scotland.}
\address{$^{9}$ UK Astronomy Technology Centre, Blackford Hill, Edinburgh EH9 3HJ, Scotland.}

\ead{i.martin@physics.gla.ac.uk}
\begin{abstract}

\noindent Thermal noise arising from mechanical dissipation in dielectric reflective coatings is expected to critically limit the
sensitivity of precision measurement systems such as high-resolution optical spectroscopy, optical frequency standards and future
generations of interferometric gravitational wave detectors. We present measurements of the effect of post-deposition heat treatment
on the temperature dependence of the mechanical dissipation in ion-beam sputtered tantalum pentoxide between 11\,K and 300\,K. We find
the temperature dependence of the dissipation is strongly dependent on the temperature at which the heat treatment was carried out,
and we have identified three dissipation peaks occurring at different heat treatment temperatures. At temperatures below 200\,K, the
magnitude of the loss was found to increase with higher heat treatment temperatures, indicating that heat treatment is a significant
factor in determining the level of coating thermal noise.

\end{abstract}

%\maketitle
\footnotetext[1]{Scottish Universities Physics Alliance}
\section{Introduction}
\label{sec:Intro}

Several long baseline gravitational wave detectors have been constructed and used in searches for gravitational radiation from
astrophysical sources \cite{Abramovici_1992,Luck_2006,Acernese_2006,Takashi_2004}. These detectors use interferometric sensing to
search for displacements, induced by gravitational waves, of test masses which are suspended as pendulums at the ends of perpendicular
arms up to 4 km in length. The test masses are coated to form mirrors which are highly reflective at 1064 nm. The thermal displacement
noise associated with the test mass mirrors is a significant limit to the sensitivity of these detectors
\cite{Saulson1990,Callen1952,Callen_Welton_1951}. The magnitude of the thermal noise is directly related to the mechanical dissipation
of the materials used to construct the mirrors. In particular, the mechanical dissipation of the amorphous dielectric mirror coatings
has been identified as an important noise source which is likely to limit the sensitivity of the generation of detectors currently
being constructed, such as Advanced LIGO \cite{Harry_AdvLIGO} and Advanced Virgo \cite{AdvVirgo_WP}, in their most sensitive frequency
band \cite{Levin1998,Nakagawa_2002_coating,Harry2002,Crooks2002}. Coating thermal noise is also expected to significantly limit the
performance of highly frequency-stabilized lasers for use in high resolution optical spectroscopy \cite{Rafac_2000}, optical frequency
standards \cite{ludlow06, Webster04} and fundamental quantum measurements \cite{Schmidt-Kaler}.

The mirror coatings used in current gravitational wave detectors are formed from alternating layers of silica (SiO$_2$) and tantalum
pentoxide (Ta$_2$O$_5$) deposited by ion-beam sputtering. Research has shown that the mechanical dissipation of these multi-layer
coatings is dominated by the dissipation of the Ta$_2$O$_5$ layers \cite{Penn2003,Crooks2004,Crooks2006} and that the total
dissipation of a multi-layer coating can be reduced by up to $\sim$ 40 $\%$ by doping the Ta$_2$O$_5$ with titania (TiO$_2$)
\cite{Harry_2007_cqg}. However, neither the process responsible for mechanical dissipation in Ta$_2$O$_5$ nor the mechanism by which
TiO$_2$ doping reduces the dissipation is as yet understood.

In a recent paper we reported data showing a low temperature dissipation peak in a Ta$_2$O$_5$ coating doped with TiO$_2$ and heat
treated at 600 $^{\circ}$C \cite{Iain_2008}. There is evidence that this peak arises from a thermally activated dissipation mechanism,
possibly related to the reorientation of Ta-O bonds within a double well potential, similar to the mechanism believed to occur in
fused silica \cite{SmallDiss,Bommel_Mason_1956,Strakna_1961}. Further work has suggested that the effect of the TiO$_2$ doping is to
increase the activation energy of the dissipation process \cite{Iain_2009}, thereby reducing the mechanical loss of the material. A
barrier height distribution function analysis revealed that the effect of TiO$_2$ doping was to shift the distribution of potential
barriers to a higher energy and to broaden the peak in the distribution. Heat treatment is another possible method of altering the
distribution of potential barrier heights. Ion-beam sputtered coatings are often heat treated after deposition to reduce the stress in
the film and to reduce the optical absorption \cite{Netterfield_2005}. Several studies have shown that post-deposition heat treatment
of ion-beam sputtered amorphous oxide layers can increase the thickness of the film, with a corresponding reduction in density
\cite{Netterfield_2005, Tilsch_1997,Brown_2004,Waldorf_1993}. Furthermore it is known heat treatment can significantly reduce the
mechanical loss of bulk fused silica \cite{Numata_2004}. This is believed to be a result of a reduction of internal stresses in the
silica, possibly altering the distribution of the potential barrier heights \cite{Lunin}. Thus investigating the effects of heat
treatment on the mechanical loss of Ta$_2$O$_5$ is of particular interest for understanding, and perhaps reducing, the mechanical
dissipation. In addition to providing insights into the mechanisms of energy dissipation in Ta$_2$O$_5$, measurements of the
temperature dependence of the coating dissipation are critical for the design of future gravitational wave detectors such as ET
\cite{ET_proposal} and LCGT \cite{kuroda10} which may use cryogenic cooling to reduce thermal noise.

\section{Sample preparation and experimental technique}
\begin{figure}[h!]
\centering
\includegraphics[width=275pt]{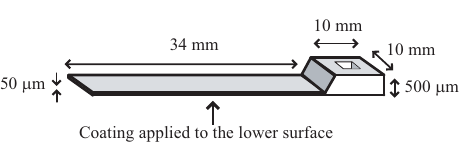}
\caption{\textit{A schematic diagram of one of the silicon cantilever samples used for the coating loss measurements. The cantilever
length is parallel to the $\langle110\rangle$ crystal axis.}} \label{fig:cantilever}
\end{figure}

The Ta$_2$O$_5$ coatings were deposited onto rectangular silicon cantilever substrates. These substrates consist of a 0.5-mm-thick
clamping block and a thinner flexing portion of length 34-mm and thickness 50\,$\upmu$m, as shown in Figure \ref{fig:cantilever}. The
sample geometry was designed to reduce frictional energy losses into the clamp used to support the cantilever during mechanical loss
measurements \cite{Quinn1995, Yasamura2000}. A thermal oxide layer of approximately 30 nm thickness was grown on the cantilevers prior
to coating to ensure proper adhesion of the Ta$_2$O$_5$. A Ta$_2$O$_5$ film (0.49 $\pm$ 0.1)\,$\upmu$m in thickness was deposited on
one face of the flexing part of each cantilever by ion-beam sputtering using argon as the sputtering ion. Four fused silica disks,
25.4 mm in diameter, were coated in the same coating run to serve as witness samples for microscopic analysis of the coating
structure. A coated cantilever, a nominally identical uncoated cantilever and a coated silica disk were heat treated in air at each of
the following temperatures: 300, 400, 600 and 800\,$^\circ$C. The samples were heated to the desired temperature at 2\,$^\circ$C per
minute and held at this temperature for 24 hours after which they were allowed to cool naturally. The coating and heat treatment
process was carried out by CSIRO\footnotemark[2]. \footnotetext[2]{Industrial Physics Division, Commonwealth Scientifc and Industrial
Research Organisation, West Lindfield, NSW, Australia}
Ion-beam sputtered Ta$_2$O$_5$ films have been observed to crystallize when heated to approximately 600 to 700\,$^\circ$C
\cite{CSIRO_report}, to the detriment of their optical properties. The highest heat treatment temperature of 800\,$^\circ$C was chosen
to allow study  of the mechanical loss on either side of this transition region.

The mechanical dissipation factors of several resonant modes of each coated cantilever and the associated uncoated control cantilever
were measured between 11 and 300\,K in a cryostat using a 'ring-down' technique, in which a resonant mode of the sample of angular
frequency $\omega_0$ was excited to an amplitude $A_0$ and the vibration allowed to decay freely. The time dependence of the amplitude
ring-down $A(t)$ is given by:
\begin{equation}A(t)=A_0e^{-\phi(\omega_0)\omega_0t/2},
\label{equ:coatingloss}
\end{equation}
where $\phi(\omega_0)$ is the mechanical dissipation factor. The cantilever was mounted in a stainless steel clamp within the vacuum
chamber of the cryostat and the bending modes excited by an electrostatic actuator located a few mm below the cantilever. The
frequency of the n$^{\textrm{\scriptsize{th}}}$ bending mode can be calculated as follows \cite{TheoryVibrations}:
\begin{equation}\omega_n=\frac{\alpha_n^2t}{4\pi\sqrt3L^2}\left(\frac{Y}{\rho}\right)^{1/2}
\label{eqn:modes}
\end{equation}
where $\rho$ is the density, $Y$ the Young's modulus, $t$ the thickness and $L$ the length of the cantilever. $\alpha_n$ is a
numerical factor which takes the values 1.875, 4.694, 7.855 and 10.996 for $n$ = 1 to 4 respectively. For $n\geq5$, $\alpha_n$ can be
approximated as $\alpha_n\simeq(2n-1)\pi/2$. The vibration amplitude was measured by means of a laser beam reflected from the
cantilever surface and directed onto a photodiode sensor outside the cryostat. The sensor consisted of a rectangular photodiode masked
so that a thin triangular strip of the surface was exposed. Vibrations of the cantilever resulted in motion of the reflected laser
beam along the triangular sensor, producing a signal proportional to the vibration amplitude. This sensor arrangement was found to be
more practical than a quadrant photodiode due to the large amplitude of motion of the laser spot.

Several measurement cycles, in each of which the sample temperature was increased incrementally from 11 to 300\,K, were carried out.
Repeated ring-down measurements at each temperature generally showed a variation in loss factor of less than 5 $\%$ for each mode. The
variation in the loss factor between the various temperature cycles was typically less than 10$\%$. The temperature of the cantilever
was measured using a silicon-diode sensor (Lakeshore DT-670 A) mounted within the clamp immediately below the fixed end of the
cantilever.

\section{Results}

\begin{figure}[h!]
\centering
\includegraphics[width=275pt]{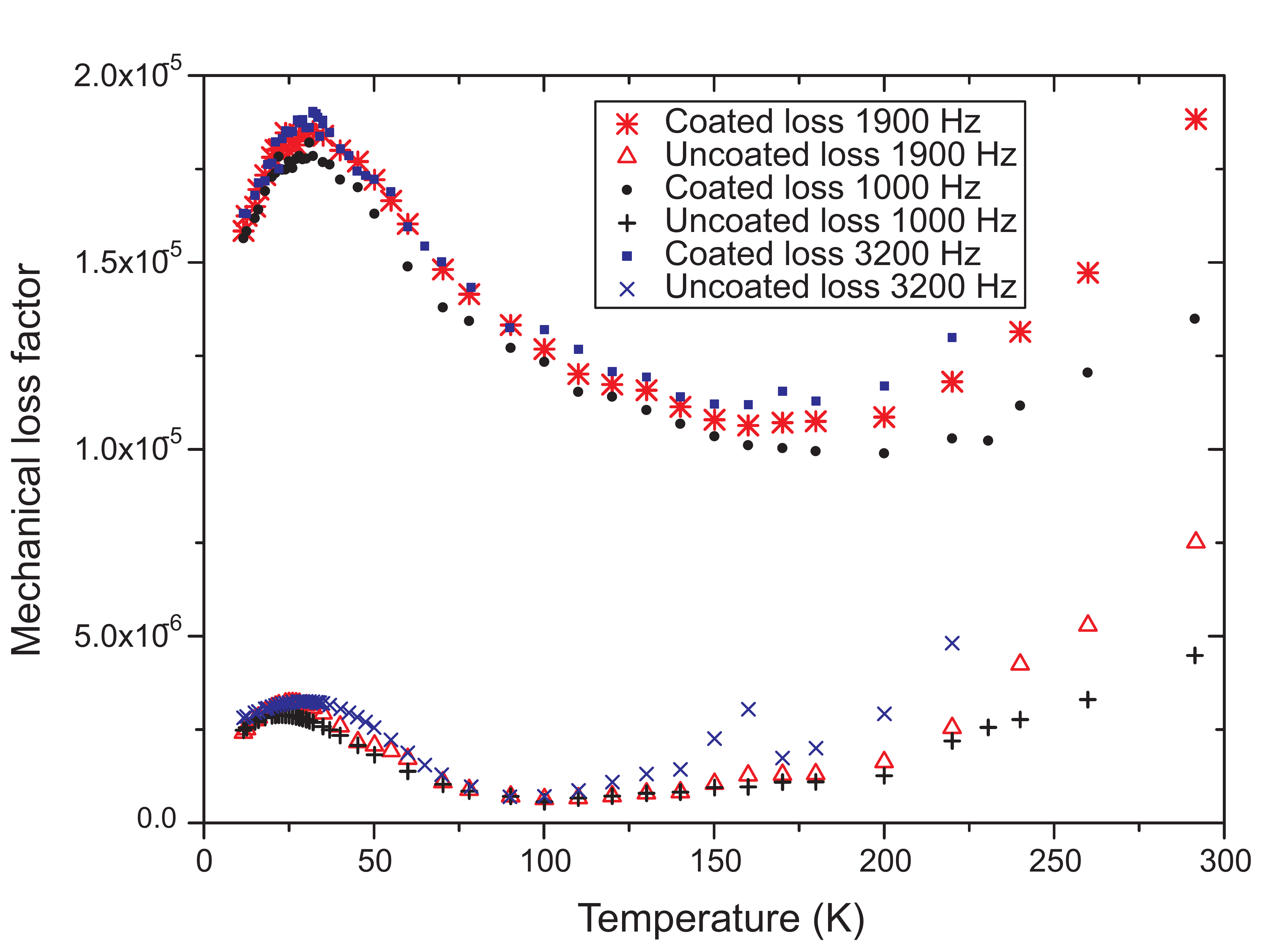}
\caption{\textit{Measured mechanical loss for a cantilever coated with Ta$_2$O$_5$ and an uncoated control cantilever heat treated at
300 $^{\circ}$C}} \label{fig:300c_measured}
\end{figure}

Figure \ref{fig:300c_measured} shows, as an example, the loss factors measured for the coated and uncoated cantilevers heat treated at
300\,$^\circ$C. The increase in loss due to the presence of the Ta$_2$O$_5$ film can be clearly observed throughout the temperature
range studied. Of particular note is the peak in the loss of the coated cantilever at approximately 35\,K - this will be discussed in
more detail in the following section. The dissipation of the uncoated cantilever also exhibits a peak at between 25 to 30\,K. A
similar peak was observed for all of the uncoated samples studied. Peaks have been observed at similar temperatures in silica films
grown on silicon by wet thermal oxidation \cite{White_Pohl_1995}, and it seems likely therefore that these peaks are due to the
thermal oxide layer grown on all of the cantilevers.

\begin{figure}[h!]
\centering
\includegraphics[width=275pt]{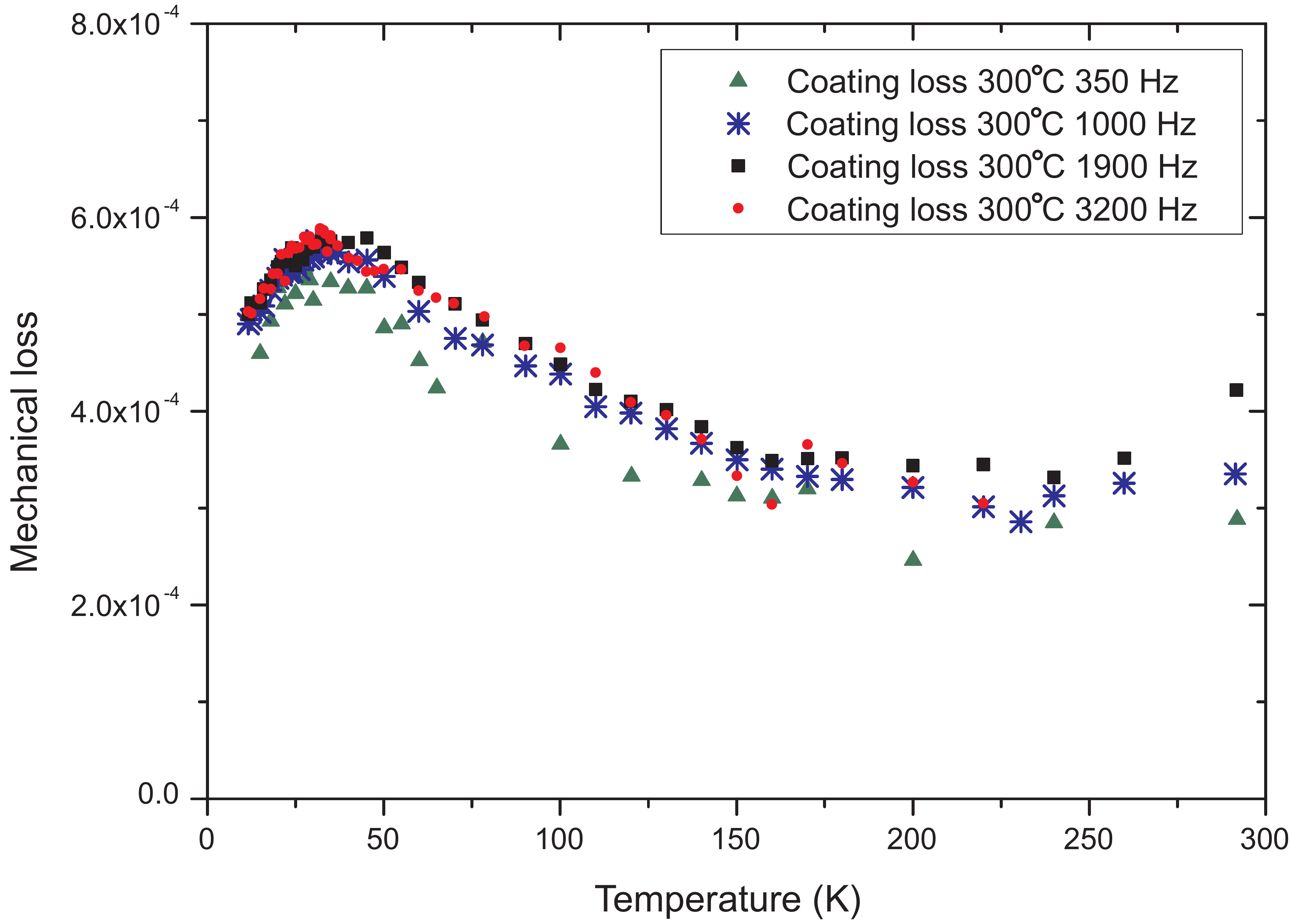}
\caption{\textit{Mechanical loss of the Ta$_2$O$_5$ coating heat treated at 300 $^{\circ}$C calculated from the measured data in
Figure \ref{fig:300c_measured}}} \label{fig:300c_coatingloss}
\end{figure}

The mechanical loss of the Ta$_2$O$_5$ film was calculated from the difference in the loss factors of the coated and uncoated
cantilevers using the following equation \cite{Berry_1975}:
\begin{equation}
\phi(\omega_0)_{\textrm{\scriptsize{coating}}}=\frac{Y_{\textrm{\scriptsize{s}}}t_{\textrm{\scriptsize{s}}}}{3Y_{\textrm{\scriptsize{c}}}t_{\textrm{\scriptsize{c}}}}
(\phi(\omega_0)_{\textrm{\scriptsize{coated}}}-\phi(\omega_0)_{\textrm{\scriptsize{substrate}}}), \label{equ:coatingloss}
\end{equation}
where $\omega_0$ is the angular frequency of the mode,  $\phi(\omega_0)_{\textrm{\scriptsize{coated}}}$ is the loss factor of the
coated cantilever, $\phi(\omega_0)_{\textrm{\scriptsize{substrate}}}$ is the loss factor of the cantilever prior to coating,
$t_{\textrm{\scriptsize{s}}}$ and $Y_{\textrm{\scriptsize{s}}}$ are the thickness and Young's modulus of the substrate respectively
and $t_{\textrm{\scriptsize{c}}}$ and $Y_{\textrm{\scriptsize{c}}}$ are respectively the thickness and Young's modulus of the coating.
The Young's moduli of silicon and Ta$_2$O$_5$ were taken to be 166 GPa \cite{Touloukian70} and (140 $\pm$ 15) GPa
\cite{tantala_modulus} respectively. Assuming that the temperature dependence of the Young's modulus of Ta$_2$O$_5$ is typical of
other amorphous oxides \cite{Marx_1952}, then its effect on the calculation over the temperature range studied here is negligible.

\begin{figure}[h!]
\centering
\includegraphics[width=275pt]{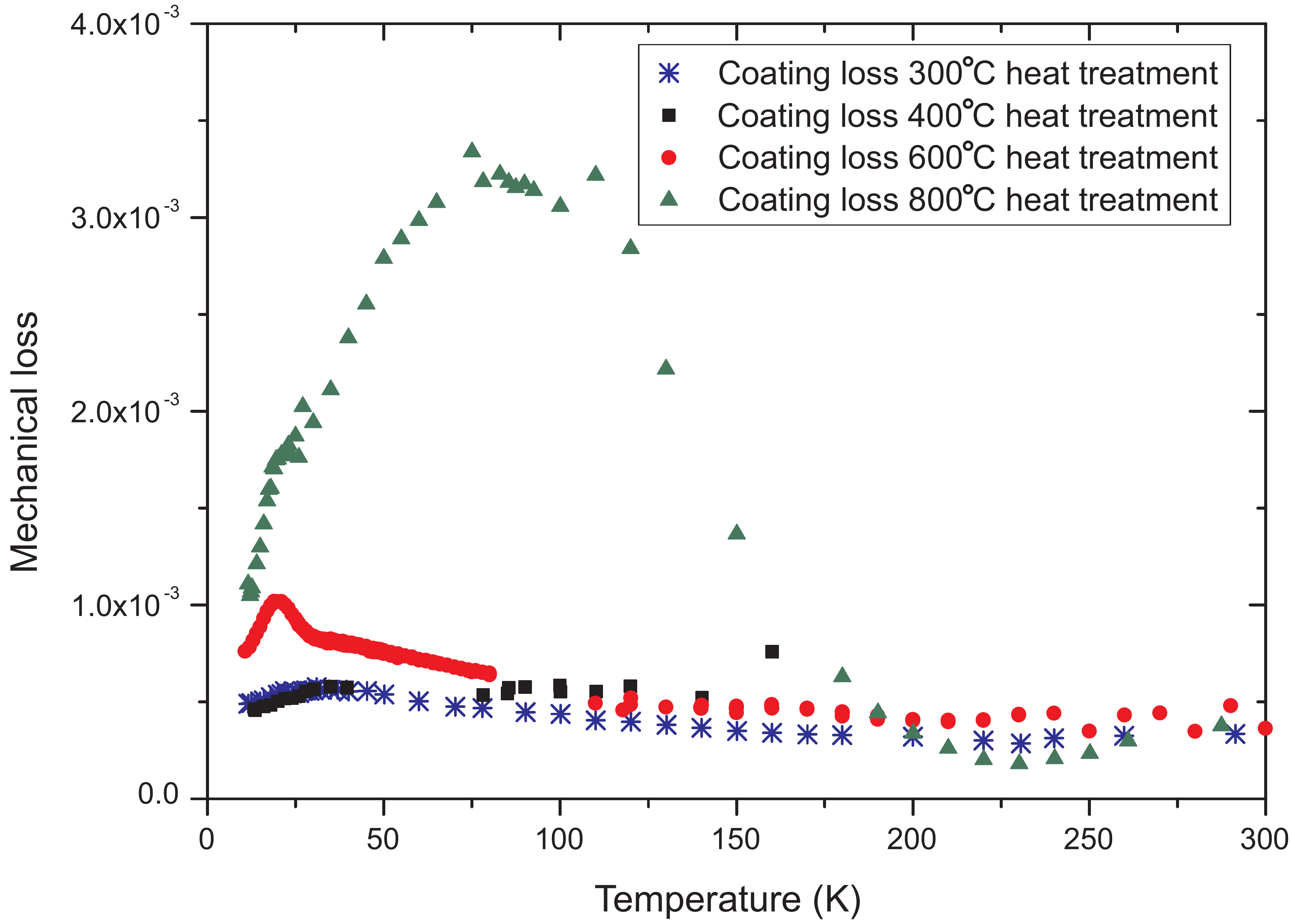}
\caption{\textit{The calculated mechanical loss factors of the various Ta$_2$O$_5$ coatings at approximately 1000 Hz}}
\label{fig:coatingloss_1000Hz}
\end{figure}

\begin{figure}[h!]
\centering
\includegraphics[width=275pt]{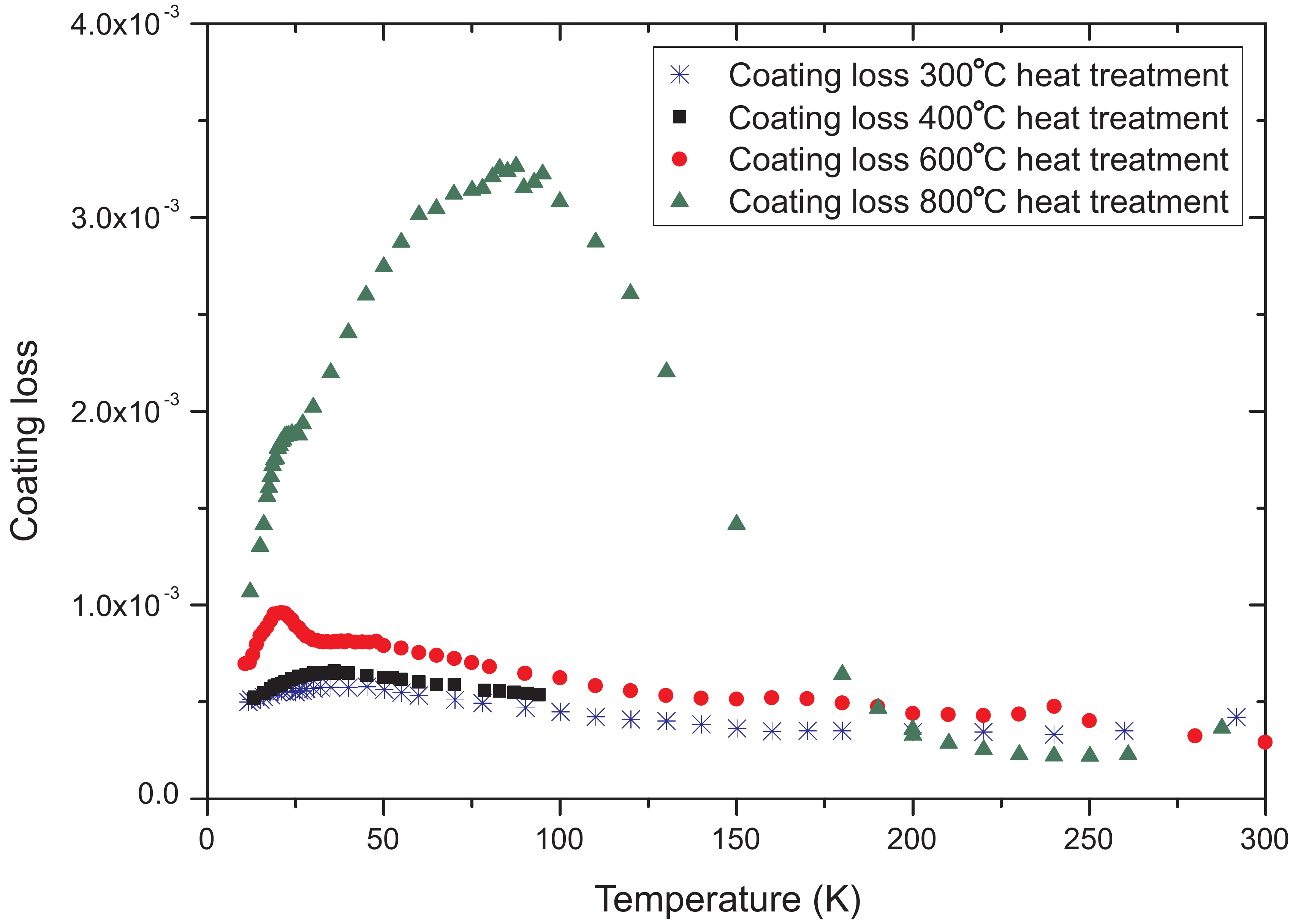}
\caption{\textit{The calculated mechanical loss factors of the various Ta$_2$O$_5$ coatings at approximately 1900 Hz}}
\label{fig:coatingloss_1900Hz}
\end{figure}

\begin{figure}[h!]
\centering
\includegraphics[width=275pt]{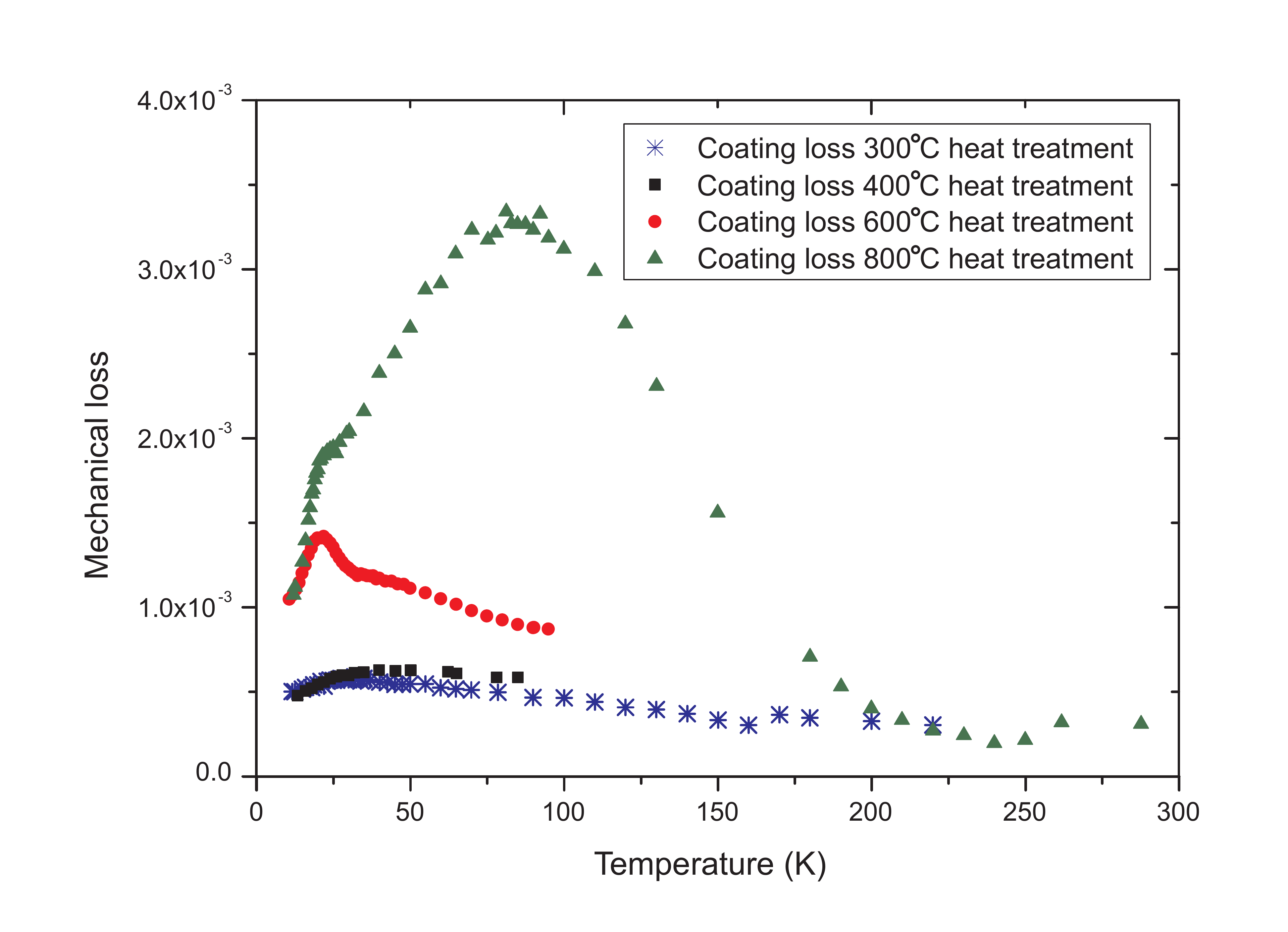}
\caption{\textit{The calculated mechanical loss factors of the various Ta$_2$O$_5$ coatings at approximately 3000 Hz}}
\label{fig:coatingloss_3000Hz}
\end{figure}

Figure \ref{fig:300c_coatingloss} shows the calculated mechanical dissipation of the Ta$_2$O$_5$ film heat treated at 300\,$^\circ$C,
obtained from the data in Figure \ref{fig:300c_measured} using Equation \ref{equ:coatingloss}. The same procedure was followed to find
the loss of the coatings which had been heat treated at 400, 600 and 800$^\circ$C. Data  was obtained at several mode frequencies for
each sample, and Figures \ref{fig:coatingloss_1000Hz} to \ref{fig:coatingloss_3000Hz} show comparisons of the loss of the series of
coatings at 1000 Hz, 1900 Hz, 3000 Hz and 350 Hz respectively.

\begin{figure}[h!]
\centering
\includegraphics[width=275pt]{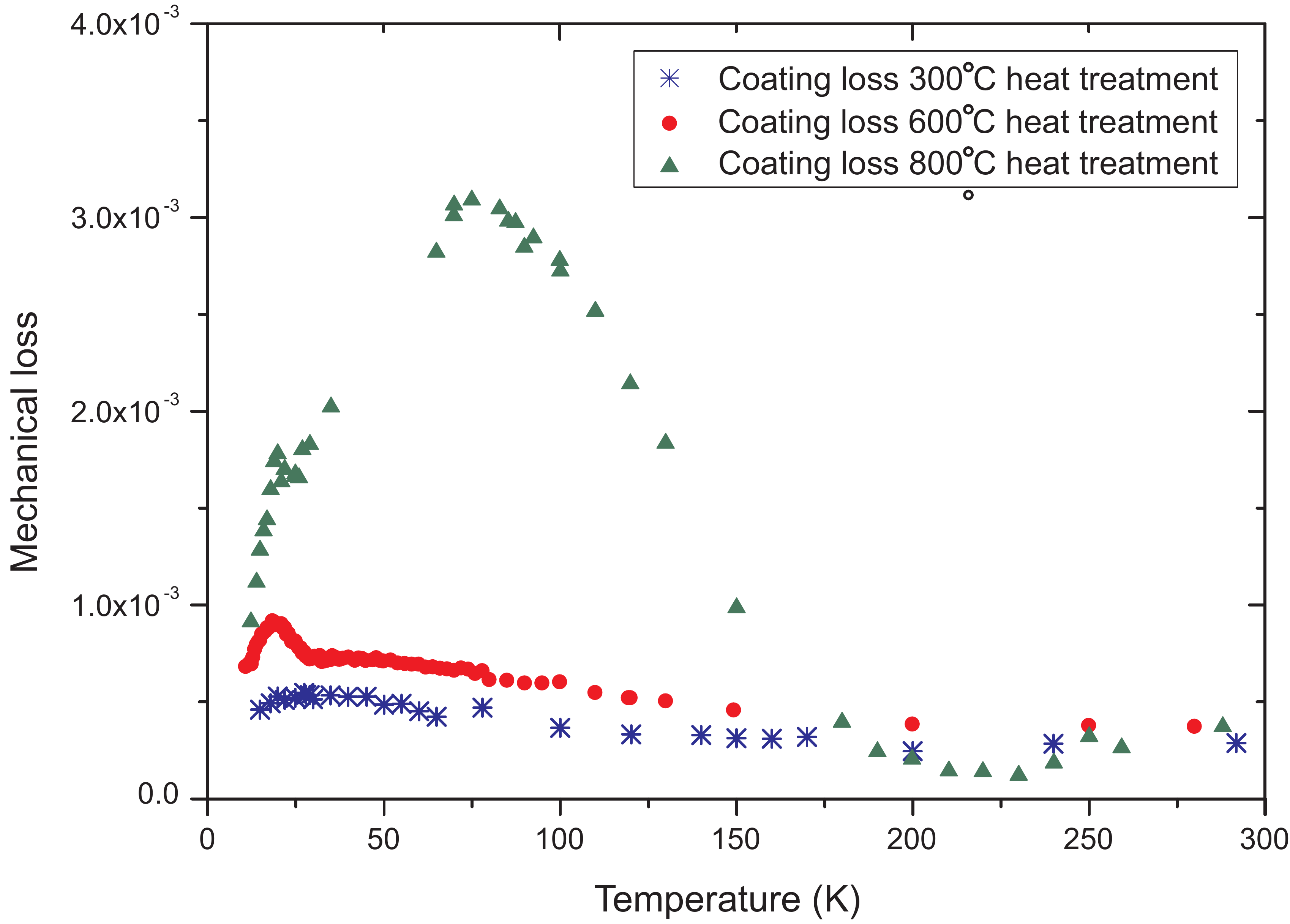}
\caption{\textit{The calculated mechanical loss factors of the various Ta$_2$O$_5$ coatings at approximately 350 Hz}}
\label{fig:coatingloss_350Hz}
\end{figure}

The results clearly show that the temperature of the post-deposition heat treatment had a significant effect on the temperature
dependence of the mechanical loss factor of the Ta$_2$O$_5$ coatings. There are several notable features in the data. Firstly, a broad
peak in the dissipation of the Ta$_2$O$_5$ film heat treated at 300\,$^\circ$C was observed at approximately 35\,K. The data for the
coated cantilever which was heat treated at 400\,$^\circ$C is incomplete as the cantilever cracked in the clamp during a measurement
cycle. The breakage was indicated by a sudden change in the resonant frequencies of the cantilever and an increase in the measured
mechanical dissipation factors. However, below 100\,K the loss of this coating follows a similar trend to the coating heat treated at
300\,$^\circ$C, with a broad dissipation peak at 35\,K.

Secondly, a sharper additional peak in the dissipation of the Ta$_2$O$_5$ film heat treated at 600\,$^\circ$C was observed at
approximately 20\,K. This is in qualitative agreement with previous measurements of Ta$_2$O$_5$ films supplied by LMA\footnotemark[3]
\footnotetext[3]{Laboratoire des Mat\'{e}riaux Avanc\'{e}s, LMA, CNRS-IN2P3, France, http://lma.in2p3.fr.} and heat treated at the
same temperature \cite{Iain_2008, Iain_2009}. This peak at 20\,K is significantly narrower and higher than the peak at 35\,K observed
in the coatings heat treated at 300 and 400\,$^\circ$C. At temperatures above 40\,K, the dissipation of the coating heat treated at
600\,$^\circ$C shows a similar trend in shape to that for the 300 and 400\,$^\circ$C coatings, although the absolute level of
dissipation is somewhat higher below 250\,K. This suggests that the 35\,K peak may also be present in the film heat-treated at
600\,$^\circ$C, partially underlying the tail of the 20\,K peak.

Thirdly, the Ta$_2$O$_5$ coating treated at 800\,$^\circ$C exhibits a large and very broad dissipation peak at 80 to 90\,K. A small
plateau in the low temperature edge of this peak is apparent at each measurement frequency, suggesting that the 20\,K dissipation peak
is also present. At temperatures between 200 and 250\,K, this coating has a lower dissipation than those heat treated at lower
temperatures, with the minimum dissipation occurring at approximately 225\,K. The increase in the dissipation between 225\,K and
300\,K may be indicative of an additional dissipation peak occurring above room temperature. However, further investigation was beyond
the scope of the current study, as the sensitivity with which the loss factor of the coating can be measured is significantly reduced
at room temperature and above due to the relatively high dissipation factor of the silicon cantilever substrates at these
temperatures.

\begin{figure}[h!]
\centering
\includegraphics[width=300pt]{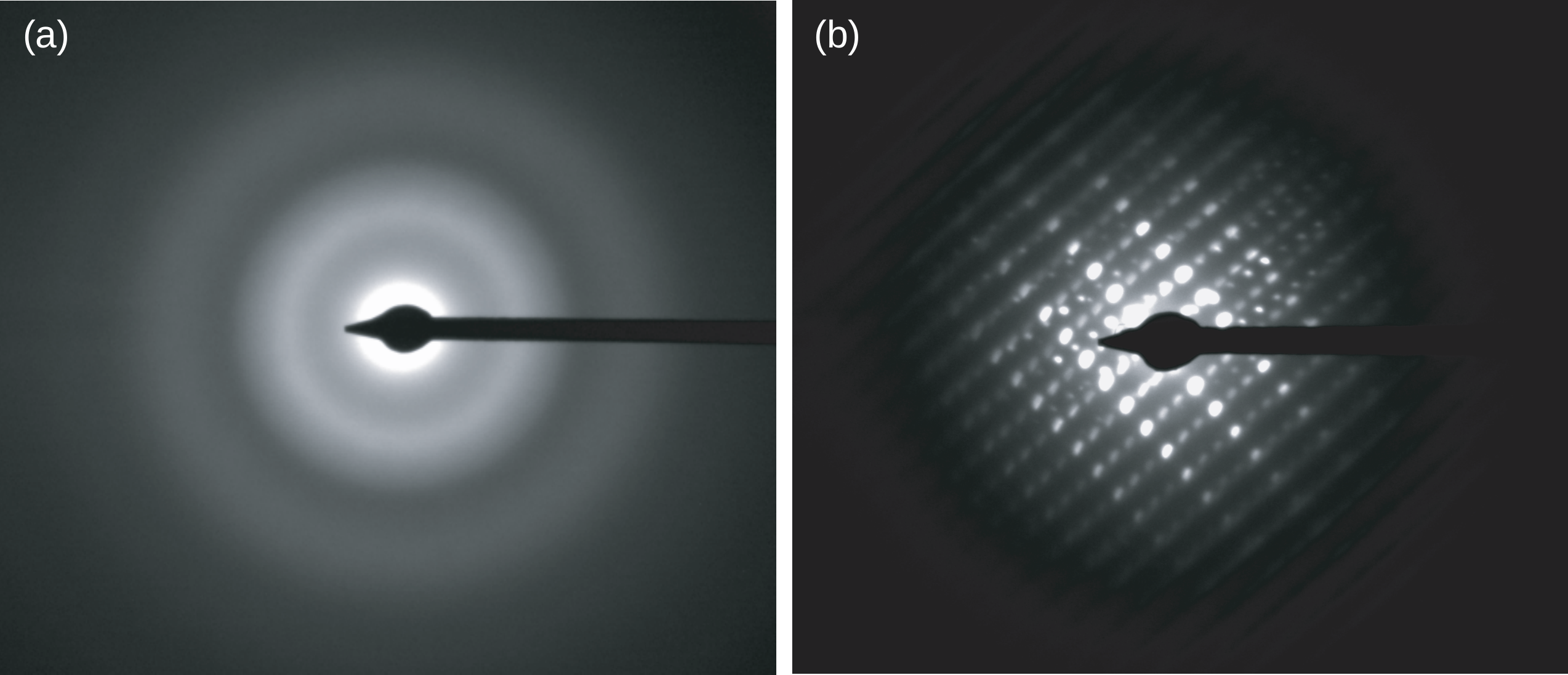}
\caption{\textit{(a)Electron diffraction pattern from the Ta$_2$O$_5$ coating heat treated at 600 $^{\circ}$C. The diffuse rings
indicate an amorphous structure. (b) Electron diffraction pattern from the Ta$_2$O$_5$ coating heat treated at 800 $^{\circ}$C,
indicating crystalline structure.}} \label{fig:e_diff}
\end{figure}
\begin{figure}[h!]
\centering
\includegraphics[width=300 pt]{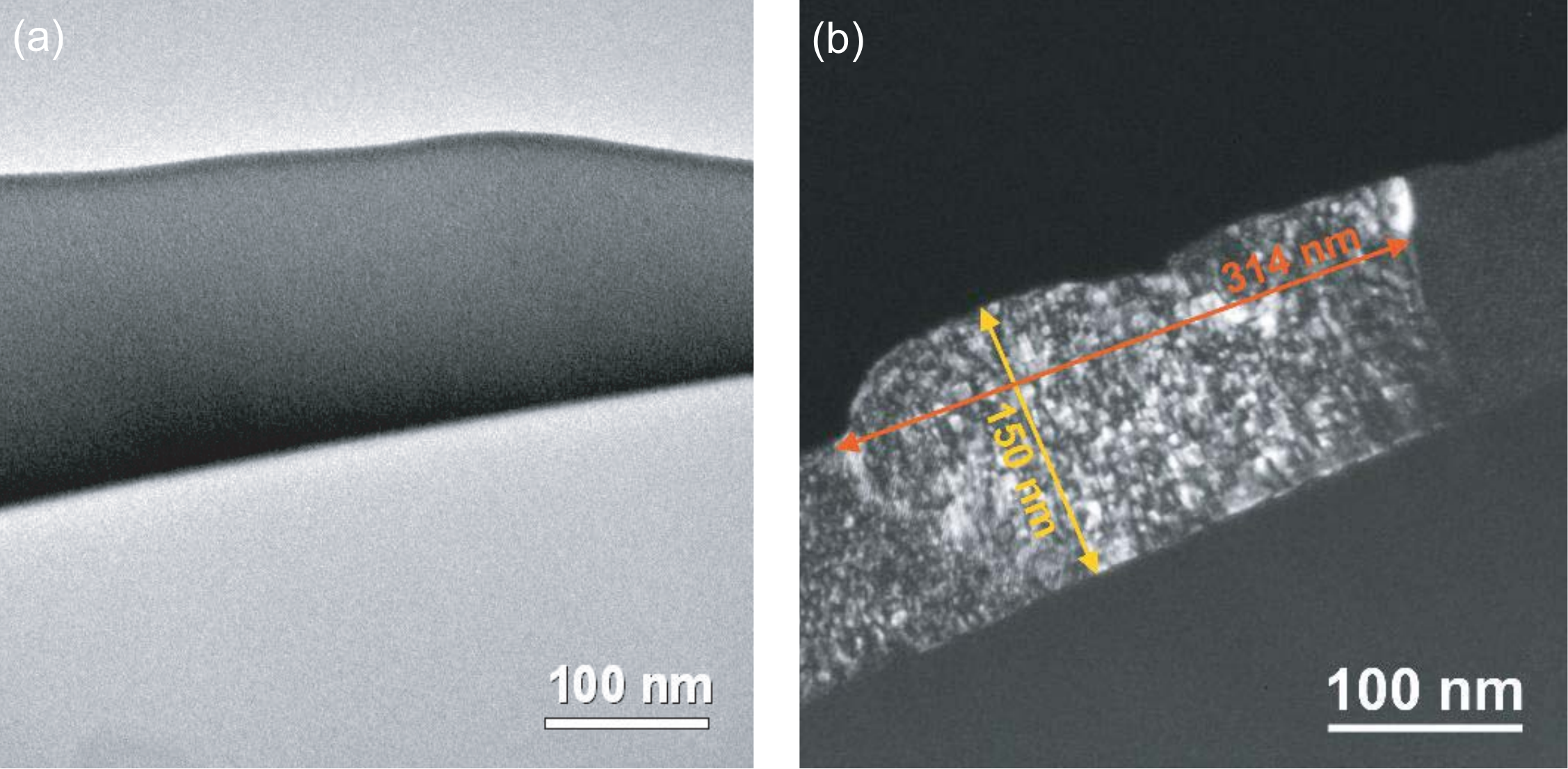}
\caption{\textit{(a) Bright field TEM image of a typical amorphous Ta$_2$O$_5$ film (darker region), heat treated at
600\,$^\circ$C.(b) Dark field TEM image of a crystallized Ta$_2$O$_5$ film (lighter region) which was heat treated at
800$^{\circ}$C.}} \label{fig:800_tantala_image}
\end{figure}

As noted above, ion-beam sputtered Ta$_2$O$_5$ has been observed to crystallize at temperatures between 600 and 700\,$^\circ$C. The
structure of the Ta$_2$O$_5$ films was investigated by electron diffraction measurements of the the witness samples which were coated
and heat treated at the same time as the cantilevers. The results, shown in Figure \ref{fig:e_diff}, indicate that the coatings heat
treated at 600\,$^\circ$C and below are amorphous, while the coating heat treated at 800\,$^\circ$C has a crystalline structure. The
difference in structure is also apparent from comparison of transmission electron microscope (TEM) images of the coatings heat treated
at 600 and 800\,$^\circ$C, as shown in Figure \ref{fig:800_tantala_image}. It is likely therefore that the large dissipation peak
centered on 80 to 90\,K is associated with the poly-crystalline structure of the Ta$_2$O$_5$ induced by the heat treatment at
800\,$^\circ$C. One possible dissipation mechanism in poly-crystalline materials is phonon scattering at the grain boundaries
\cite{NowickandBerry}.

\section{Analysis}

\begin{figure}[h!]
\centering
\includegraphics[width=250pt]{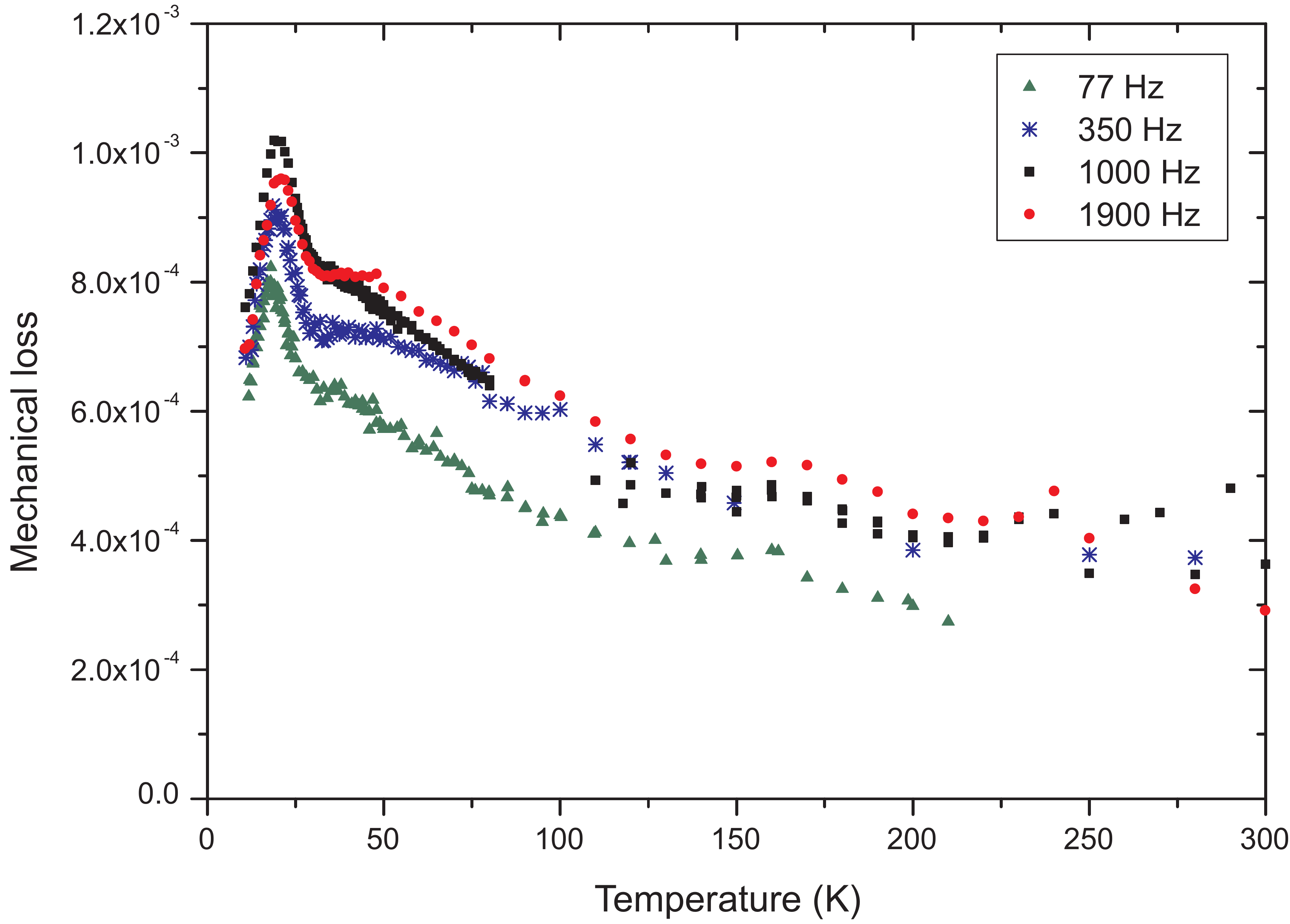}
\caption{\textit{Mechanical dissipation of the coating heat treated at 600 $^{\circ}$C at several frequencies, showing the frequency
dependence of the temperature of the peak in the low temperature dissipation.}} \label{fig:600C_coating_loss}
\end{figure}

For both the dissipation peak near 20\,K and the peak near 35\,K, the temperature at which the peak occurred was found to vary with
frequency. This behaviour, which can be seen clearly in Figure \ref{fig:600C_coating_loss} for the coating heat treated at
600\,$^\circ$C, is a typical characteristic of a thermally activated dissipation process. These processes can be characterized by a
rate constant, $\tau_0$, and an activation energy, $E_{\textrm{\scriptsize{a}}}$, which are related by the Arrhenius equation
\cite{NowickandBerry}:
\begin{equation}
\tau=\tau_0e^{E_\textrm{\scriptsize{a}}/k_BT} \label{eqn:Arrhenius}
\end{equation}
where $\tau$ is the relaxation time associated with the dissipative system returning to equilibrium after a perturbation. Analysis of
these dissipation processes shows that the temperature of the dissipation peak, $T_{\textrm{\scriptsize{{peak}}}}$, at angular
frequency $\omega_0$ is related to the activation energy and rate constant as follows \cite{NowickandBerry}:
\begin{equation}
\omega_0\tau_0e^{E_\textrm{\scriptsize{a}}/k_BT_{\textrm{\scriptsize{peak}}}}=1.
\end{equation}

\begin{figure}[h!]\centering
\includegraphics[width=250 pt]{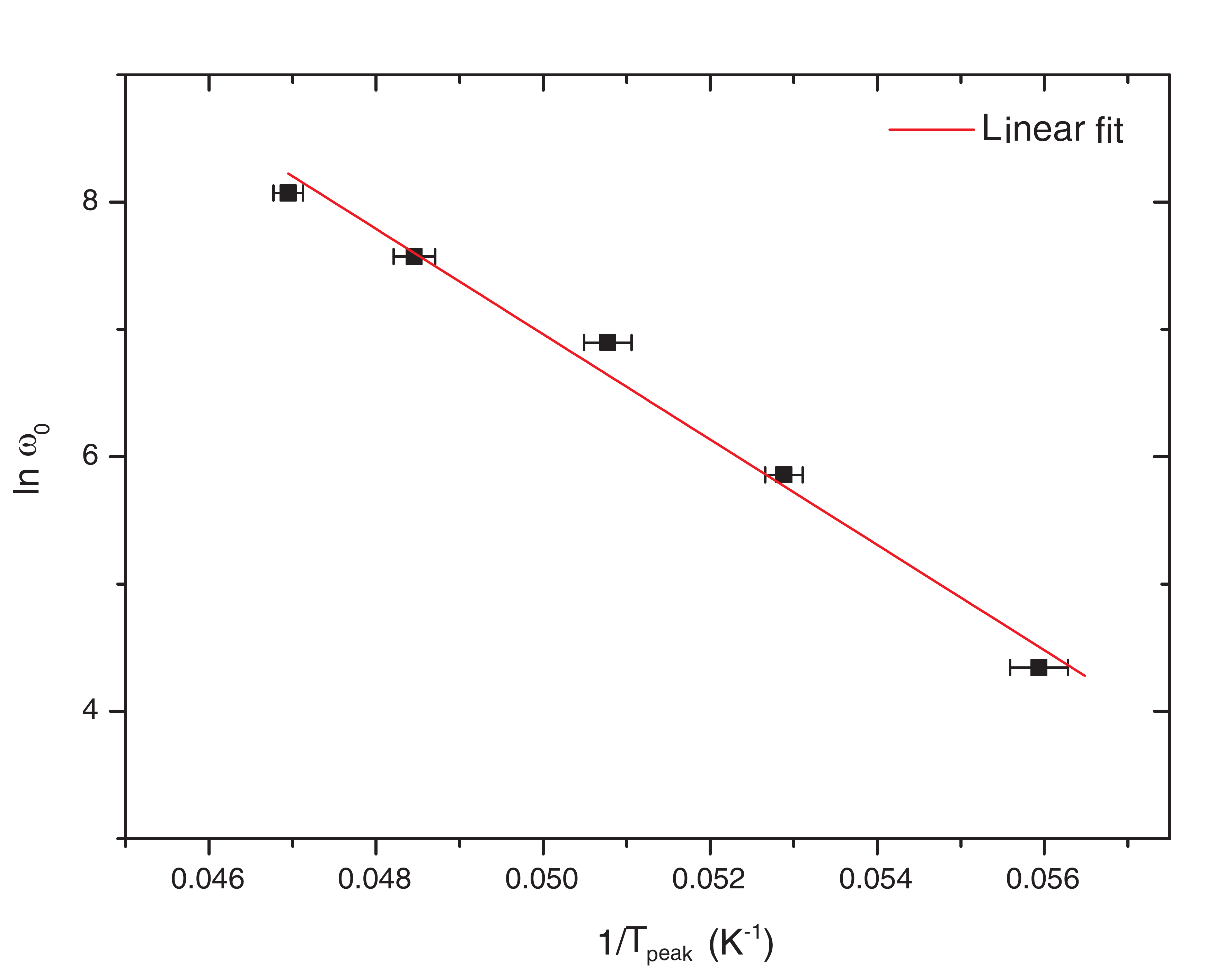}
\caption{\textit{Arrhenius plot for loss peak at 20\,K observed in the Ta$_2$O$_5$ coating heat treated at 600 $^{\circ}$C}}
\label{fig:arr_600C}
\end{figure}

Thus a plot of $\log\omega_0$ against $1/T_{\textrm{\scriptsize{{peak}}}}$ should yield a straight line, from which the activation
energy and rate constant for the dissipation process can be obtained. Figure \ref{fig:arr_600C} shows this analysis for the peak near
20 \,K in the coating heat treated at 600\,$^\circ$C. The activation energy and rate constant, calculated from the linear fit, were
found to be (35.6 $\pm$ 2.5) meV and (9.9 $\pm$ 0.5)$\times 10^{-13}$ s respectively. The activation energy is somewhat higher than a
previous measurement of (28.6 $\pm$ 1.6) meV for Ta$_2$O$_5$ deposited by LMA. The difference in activation energy could arise from
differences in the detailed structure of the amorphous coating induced by variations in the ion-beam sputtering deposition process
between coating vendors.

\begin{figure}[h!]
\centering
\includegraphics[width=220 pt]{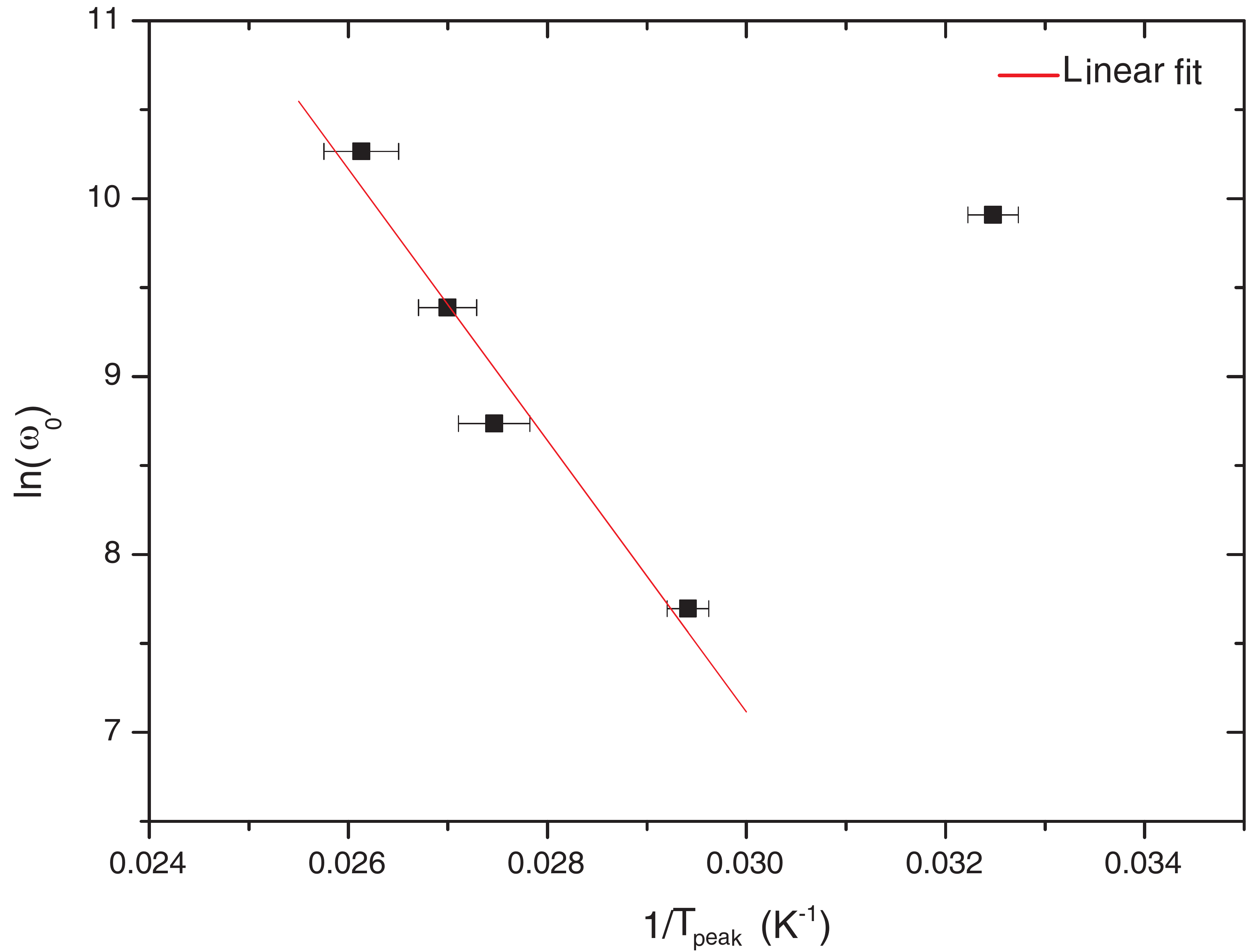}
\caption{\textit{Arrhenius plot for loss peak at 35\,K observed in the Ta$_2$O$_5$ coating heat treated at 300 $^{\circ}$C}}
\label{fig:arr_300_C}
\end{figure}

The Arrhenius plot for the peak near 35\,K observed for the coating heat treated at 300\,$^\circ$C is shown in Figure
\ref{fig:arr_300_C}. A straight line could be fitted to the data for four of the modes studied. However, the point on the Arrhenius
plot corresponding to the mode at 3200 Hz clearly does not follow this linear trend. It is possible that a torsional mode close in
frequency to the 5$^{\textrm{th}}$ bending mode was measured unintentionally, which may explain the difference in the loss peak for
this mode. If this point is neglected the fit for the remaining modes gives an activation energy of (66 $\pm$ 10) meV and a rate
constant of (9.4 $\pm$ 0.9)$\times 10^{-14}$ s.

In many amorphous solids, the dissipation at temperatures above 10\,K is thought to result from thermally activated reorientations of
bonds over potential barriers of height $V$ \cite{Gilroy_Phillips_1981}. If the dissipation is modelled as arising from transitions in
an asymmetric double-well potential in some local configuration coordinate, it can be shown that the measured mechanical loss is
related to the distribution of energy barrier heights $g(V)$ by the following expression \cite{Cahill_Topp_1996}:
\begin{equation}
\phi = \frac{\pi\gamma^2f_0}{C_{\textrm{\scriptsize{ii}}}}k_\textrm{\scriptsize{B}}Tg(V), \label{eqn:Topp_loss}
\end{equation}
where
\begin{equation}
V = k_\textrm{\scriptsize{B}}T\textrm{ln}(\frac{1}{\omega\tau_0}). \label{eqn:Topp_loss_V}
\end{equation}
and $C_{\textrm{\scriptsize{ii}}}$ is the appropriate elastic constant for the mode-shape under consideration, $\gamma$ is the elastic
coupling constant which represents the coupling between the defect (e.g. a bond re-orienting within the double well potential) and the
applied strain and $f_0$ is a constant representing the asymmetry between the depth of the two potential wells in each double-well
system. The relaxation time $\tau$ associated with a barrier height $V$ is given by Equation \ref{eqn:Arrhenius}. The assumptions made
in this model are discussed in detail by Topp and Cahill \cite{Cahill_Topp_1996}.

\begin{figure}[h!]
\centering
\includegraphics[width=225 pt]{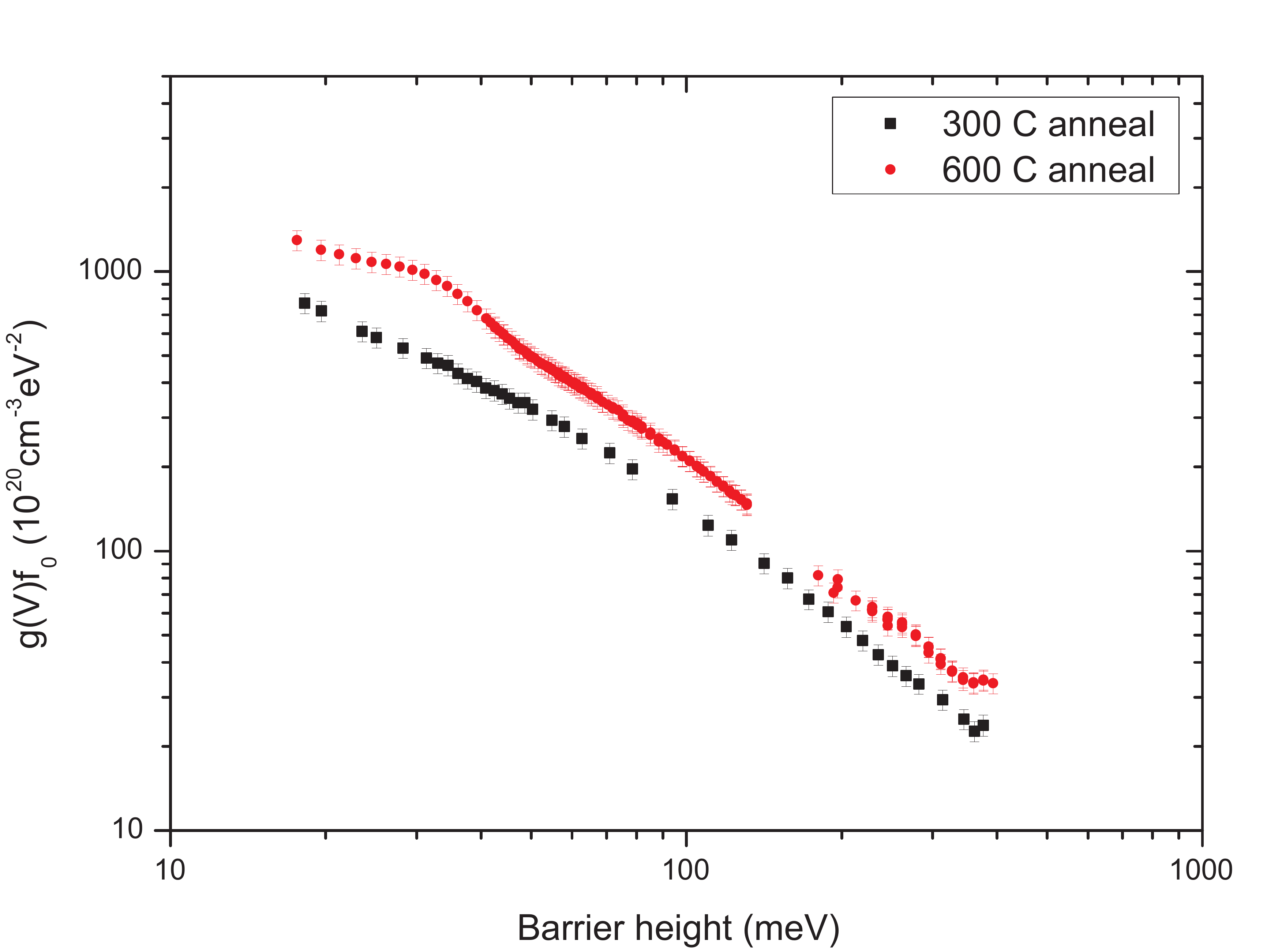}
\caption{\textit{Comparison of the calculated potential barrier height distribution function $g(V)f_0$ as a function of barrier height
for Ta$_2$O$_5$ heat treated at 300\,$^\circ$C and 600\,$^\circ$C, using the data measured at $\sim$ 1000 Hz.}} \label{fig:barriers}
\end{figure}

Thus the measured mechanical loss can be used to extract the distribution of barrier heights (i.e.\,activation energies) for the loss
mechanism associated with a particular dissipation peak. The extracted barrier height distributions for the Ta$_2$O$_5$ heat treated
at 300\,$^\circ$C and 600\,$^\circ$C are shown in Figure \ref{fig:barriers}. At all barrier energies, the magnitude of the function
$g(V)f_0$ is larger for the coating heat treated at 600\,$^\circ$C. An interpretation of this is that there are more barriers
contributing to the dissipation in this coating, resulting in a higher mechanical dissipation. The feature in the distribution
function at a barrier height energy of 35 meV corresponds to the average activation energy calculated from the Arrhenius plot. This
feature appears to be superimposed on a background distribution similar to that found for the coating heat treated at 300\,$^\circ$C.

The shape of the barrier height distribution for the 35\,K loss peak in the coating heat treated at 300\,$^\circ$C is broadly similar
to the distribution found in bulk fused silica \cite{Cahill_Topp_1996}, with $g(V)f_0$ decreasing smoothly with increasing barrier
energy. We postulate that the dissipation mechanisms in Ta$_2$O$_5$ may be similar to that in fused silica and other amorphous solids,
and be related to the reorientation of oxygen atoms, or groups of tantalum and oxygen atoms, in a double-well potential. Several
models of the dissipation mechanism in fused silica have been proposed, including motion of Si-O bonds between stable angles
\cite{Bommel_Anderson}, elongation of Si-O bonds \cite{Strakna_1961}, rotation of SiO$_4$ tetrahedral structures
\cite{Vukcevich_1972,Buchenau_1984} or of groups of these structures \cite{Reinisch_2005}. If similar processes are responsible for
the loss in Ta$_2$O$_5$, it is possible that the peaks observed at 20\,K and 35\,K correspond to two different types of motion, one of
which is only activated by changes in the local ordering induced by heat treatment at temperatures close the the crystallization
temperature. Further detailed studies of the short range structure in Ta$_2$O$_5$ coatings using electron diffraction data and
computer modelling are ongoing \cite{Riccardo_EMAG}.

\section{Conclusions}
\label{sec:conclusions}

The temperature dependence of the mechanical dissipation in Ta$_2$O$_5$ films has been found to be strongly dependent on the
temperature at which post-deposition heat treatment is carried out. We have identified three dissipation peaks arising from
dissipation mechanisms which occur after heat treatment at different temperatures. The previously observed dissipation peak at 20\,K
has been shown to arise from heat treatment close to the temperature at which the amorphous Ta$_2$O$_5$ crystallizes. The dissipation
mechanism responsible for this peak may therefore be related to pre-crystallization changes in the local ordering of the Ta$_2$O$_5$
film. A dissipation peak at approximately 35\,K was observed in Ta$_2$O$_5$ films heat-treated at 300 and 400\,$^{\circ}$C. There is
evidence that this peak is also present in Ta$_2$O$_5$ heat treated at 600\,$^\circ$C, underlying the larger dissipation peak at
20\,K. We postulate that the two low temperature dissipation peaks may arise from two different types of thermally activated
transitions. The Ta$_2$O$_5$ film heat-treated at 800$^\circ$C, which was found to have crystallized, had a very broad dissipation
peak centered on 90 - 100\,K.

Below 200\,K the dissipation of the Ta$_2$O$_5$ coatings was found to increase at higher heat treatment temperatures, suggesting that
minimizing the temperature at which post-deposition heat treatment is carried out may allow a significant decrease in the coating
thermal noise in future cryogenic gravitational wave detectors, thus extending their astronomical reach. As heat treatment is also
known to affect the optical absorption of dielectric coating, studies of the optical properties of Ta$_2$O$_5$ heat-treated at various
temperatures would be of interest.

The dissipation of Ta$_2$O$_5$ coatings at room temperature appears to be related to the broad tail of the dissipation peak at 35\,K.
However there is some evidence, particularly for the coating heat-treated at 800\,$^\circ$C, of an increase in dissipation close to
room temperature. Thus further study of the dissipation in Ta$_2$O$_5$ at temperatures above 300\,K would be of significant interest
to determine if additional dissipation peaks occurring at temperatures greater than 300\,K may contribute to the room temperature
dissipation.

Further work is underway to investigate the relationship between the short range structure of Ta$_2$O$_5$ films and the mechanical
loss. In addition, it would be of significant interest to study the emergence of the various dissipation peaks in Ta$_2$O$_5$ by
studying coatings heat treated at more temperature intervals between 400 and 800\,$^{\circ}$C.

\begin{ack}
We are grateful for the financial support provided by STFC, the University of Glasgow, EPSRC, the German Science Foundation under
contract SFB Transregio 7 and the National Science Foundation under grants PHY-07 57896 and PHY-05 02641 (Stanford) and NSF-0653590
(HWS). I. Martin and S. Reid are supported by an STFC Postdoctoral Fellowship and a Royal Society of Edinburgh Research Fellowship
respectively. R. Nawrodt acknowledges the support of the FP7 EU Einstein Telescope design study project under contract number 211743.
LIGO was constructed by the California Institute of Technology and Massachusetts Institute of Technology with funding from the
National Science Foundation and operates under cooperative agreement PHY-0107417. We would like to thank our colleagues in the LSC and
VIRGO collaborations and within SUPA for their interest in this work. This paper has LIGO Document Number LIGO-P1000074.
\end{ack}
\\
\footnotesize
\bibliographystyle{elsart-num}
\bibliography{template-bib}

\end{document}